\documentclass[
a4paper,%
12pt,%
]{article}

\usepackage[dvips]{graphicx}

\title{About evolution of particle transitions from one well
to another in double-well potential}

\author{Vladislav~S.~Olkhovsky and Sergei~P.~Maydanyuk\\
\em Institute for Nuclear Research, \\
\em 47, prosp.~Nauki, Kiev, 03680, Ukraine \\
\em maidan@kinr.kiev.ua
}

\date{}

\begin{document}

\maketitle

\begin{abstract}
The Poincare's period of particle oscillations between wells is
obtained in double-well potential. The dependencies of oscillation
period on transmission coefficient on distance between levels
are obtained. The cases of squared potentials and some potentials
having rounded off forms are considered specifically.
\end{abstract}


PACS number(s): 03.65.Bz; 03.65.Ge.

\section{Introduction}
\setcounter{equation}{0}

The tunneling of particle through potential barrier is an essentially
quantum phenomenon. This process involves propagation of a particle
through a classically inaccessible region. The complete information
about tunneling of particle from solution of Schr\"{o}dinger equation
with appropriate boundary conditions can be obtained in all regions.
But in practice, the exact solution of Schr\"{o}dinger equation can
be found for some simplest forms of potentials and it is difficult
enough to obtain the exact solution for arbitrary potential form.
For this reason the approximation methods are used for finding
solutions for potentials of specific form. But exact solutions, which
were found, are of a great importance, because they allow to analyze
the tunneling process in general.

It is enough difficult to obtain solutions for multi-dimension potential
forms. Therefore, in present work only one-dimensional case is considered,
for which exact solutions are obtained for some simple forms of potential
having two wells separated by barrier. Every specific form of potential
in comparison with another ones requires use of specific approach for
solving the problem, allowing some features of tunneling to be more
pronounced, while the others will be more unnoticeable. Having analyzed
the use of boundary conditions for solving the problem, we proposed
to divide various shapes of double-well potentials into two classes:
squared potentials and potentials having rounded off forms. Squared
double-well potentials have exact analytic solutions (which can be
expressed through elementary functions). Rounded off double-well
potentials have solutions expressed through special functions (if
these solutions exist). In present work after qualitative analysis
of energy levels of various forms of double-well potential the problem
of squared potential and the problem of rounded off potential (the
Morse's potential) are considered separately in next two sections.

The period of particle oscillations between two wells is one of the
most important parameters which characterize the process of tunneling.
We can obtain the value of a period on the basis of energy levels of
system. For this at the beginning we will confine ourselves to class
of systems, for which the distances between energy levels have the
exactly determined largest common devizor $\Delta$ and the following
condition is fulfilled:

\begin{equation}
E_{n} = E_{0} + \Delta*l_{n},
\end{equation}

where $l_{n} \in 0,N$. In general case in region of discrete
energy spectrum the states of these systems are described by wave
packet as follows \cite{Shiff.1957}

\begin{equation}
\psi(t) = \sum_{n} g_{n} \varphi_{n}(x) e^{-i(E_{n}-E_{0})t/\hbar} =
          \sum_{n} g_{n} \varphi_{n}(x) e^{-i\Delta l_{n} t/\hbar}
\end{equation}                                            

where $\varphi_{n}(x)$ is orthonormal wave functions of stationary states
of system satisfying the equation $\stackrel{\land}{H} \varphi_{n}(x)
= E_{n} \varphi_{n}(x)$; $\stackrel{\land}{H}$ is Hamiltonian of system;
$\sum\limits_{n} |g_{n}|^{2} = 1$ and insignificant factor $e^{-iE_{0}t /
\hbar}$ is omitted, which is common for all terms of sum $\sum\limits_{n}$.
Let's select the moment $t = 0$ as the origin of time reference.

Let the function $\psi(t)$ be determined on region $[-\frac{\pi\hbar}{\Delta},
\frac{\pi\hbar}{\Delta}]$ and satisfy the Dirichlet's conditions
\cite{Vorobiev.1986}:
a) it can divide this region into finite number of regions, in which
the function $\psi(t)$ will be continuous, monotonic and bounded;
b) if $t_{0}$ is the discontinuity point of function $\psi(t)$, then
$\psi(t_{0}+0)$ and $\psi(t_{0}-0)$ exist. Then expression (1.2) is
expansion in $t$ of function $\psi(t)$ into Fourier series which is
converging in all points of region $[-\frac{\pi\hbar}{\Delta},
\frac{\pi\hbar}{\Delta}]$. Then the function $\psi(t)$ is periodic in
time and period of oscillations (time of Poincare's cycle) is given by

\begin{equation}
T = \frac{2\pi\hbar}{\Delta}
\end{equation}                                     

The expression (1.3) determines period of oscillations of wave packet,
if energy levels in region of discrete spectrum have exactly determined
common largest divizor $\Delta$ and condition (1.1) is fulfilled.

In general case for systems, for which the distances between energy
levels in region of discrete spectrum don't have the exactly determined
divizor $\Delta$, one can select the 'quasi-cycles' with given degree
of accuracy, for which the state of system approaches the maximum
degree the initial state after a 'quasi-cycle' time
\cite{Olkhovsky.1989.IVF}.
States of such systems are localized in a confined volume of space and
the time of Poincare's cycle (which includes the required number of
'quasi-cycles') can be determined with given degree of accuracy. To
find information about some parameters for quantum systems evolving
with time in region of discrete spectrum see
\cite{Olkhovsky.1989.IVF,Senn.1992.AJPIA,Heiss.1989.PHRVA}.

Since period of particle oscillations between wells is obtained
on the basis of energy eigenvalues, much attention is paid
to the problem of solving of eigenvalue equations. For some forms of
potential the transmission coefficient through the barrier is
found. This parameter can be obtained using two approaches: in
region of continuous spectrum for particle which is incident upon
the barrier, with asymptotic velocity $\hbar k/m$, and, when condition
of WKB approximation is fulfilled, in region of discrete spectrum
for particle which initially is located in one well and then
is tunneling to another well. In case of double-well squared
potential the comparison of these two approaches for calculating
the transmission coefficient through barrier is performed. For
symmetric double-well potentials the dependence of transmission
coefficient (which is found using one of approaches) on period
of particle oscillations between wells is analyzed.

\section{Analysis of possibility of particle tunneling through barrier}
\setcounter{equation}{0}

The qualitative estimation --- will the particle oscillate between
wells or not --- can be obtained on the basis of solution analysis of
stationary Schr\"{o}dinger equation which are found for every well.

\begin{figure}[p]
\centering
\includegraphics[width=7cm]{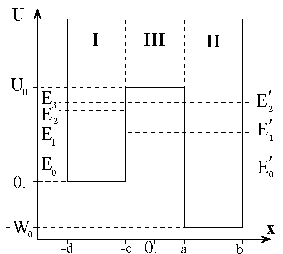}
\caption{Double-well squared potential}
\label{f1}
\end{figure}

\begin{figure}[p]
\centering
\includegraphics[width=7cm]{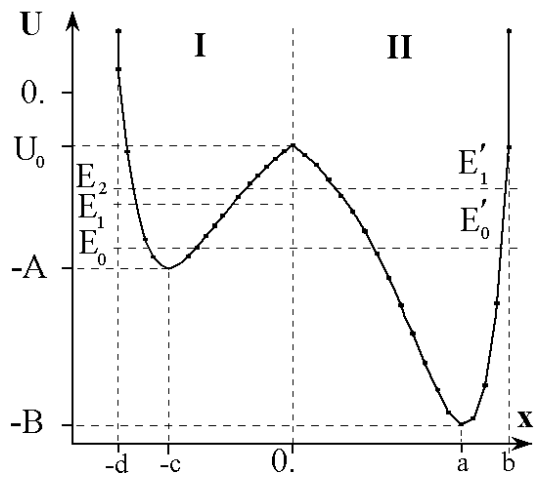}
\caption{Double-well rounded off potential}
\label{f2}
\end{figure}

Divide various forms of double-well potentials into two classes:
squared potentials (see Fig.~\ref{f1}) and potentials, which have
rounded off forms (see Fig.\ref{f2}). To obtain the energy levels,
the Schr\"{o}dinger equation is solved separately in every region
(3 regions for squared potential, see Fig.~\ref{f1}; 2 regions for
potential having rounded off form, see Fig.\ref{f2}).

As a result of this solution, the wave functions $\varphi_{i}(x)$ are
found in every region, and these functions must be continuous and
bounded (we consider discrete spectrum $E < U_{0}$) all over the region
of its determination (here, $i$ is a number of region). If wave
function is expressed through special or elementary functions,
then it will be bounded everywhere in region $-d<x<b$ (with exception of
some cases of hypergeometric functions, which must be considered
separately, see \cite{Beitman.1973,Abramowitz.1979}), and in points
$x=-d$ and $x=b$ the boundary condition is given by

\begin{equation}
\begin{array}{ll}
\varphi_{1}(-d) = 0, & \varphi_{2}(b) = 0.
\end{array}
\end{equation}                                       

This condition determines eigenvalues for energy levels as for
discrete levels. Condition of continuity of wave function and
its derivative in region $-d<x<b$ requires the equality of
solutions $\varphi(x)$ and $d\varphi(x) / dx$ for adjacent regions
in points of boundary between these regions
(in these points the discontinuity of derivative is possible).

At the beginning we consider squared potential (see Fig.~\ref{f1}).
For particle located in left well (region 1) we analyze the possible
cases of its propagation to right well in result of tunneling
through barrier. Among them we select the following cases:

1) In regions 1, 2 and 3 there are the equal levels $E_{0}$ (see
Fig.~\ref{f1}), which are found from Schr\"{o}dinger equation for
every region separately. Then particle can propagate through barrier
along level $E_{0}$. (In this case the transitions between levels
are not required for tunneling). We obtain the eigenvalue of energy
level $E_{0}$ from system (2.1) and the following system:

\begin{equation}
\begin{array}{ll}
\varphi_{1}(-c) = \varphi_{3}(-c), &
\varphi_{1}^{\prime}(-c) = \varphi_{3}^{\prime}(-c),  \\
\varphi_{3}(a) = \varphi_{2}(a), &
\varphi_{3}^{\prime}(a) = \varphi_{2}^{\prime}(a).
\end{array}
\end{equation}

2) In regions 1 and 3 there are the equal levels $E_{1}$, but
such level is absent in region 2. In regions 2 and 3 there are
the equal levels $E_{1}^{\prime}$, but such level is absent in
region 1. Then particle initially located on level $E_{1}$ in
left well, can not propagate to right well along this level. But
in region 3 wave functions corresponding to levels $E_{1}$ and
$E_{1}^{\prime}$ are not equal to zero. Therefore, in this region
the matrix element of transition from level $E_{1}$ to level
$E_{1}^{\prime}$ (and on the contrary) is not equal to zero.
Therefore, particle which initially located on level $E_{1}$ in
left well can propagate to right well with transition from level
$E_{1}$ to level $E_{1}^{\prime}$. The transition $E_{1} \to
E_{1}^{\prime}$ takes place in region of barrier 3. Eigenvalue of
level $E_{1}$ can be obtained from the system:

\begin{equation}
\begin{array}{ll}
\varphi_{1}(-c) = \varphi_{3}(-c), &
\varphi_{1}^{\prime}(-c) = \varphi_{3}^{\prime}(-c),  \\
\varphi_{1}(-d) = 0, &
\varphi_{3}(a) = 0.
\end{array}                                            
\end{equation}

In this fashion, one can find the eigenvalue for level $E_{1}^{\prime}$
from the following system:

\begin{equation}
\begin{array}{ll}
\varphi_{2}(a) = \varphi_{3}(a), &
\varphi_{2}(b) = 0,  \\
\varphi_{2}^{\prime}(a) = \varphi_{3}^{\prime}(a), &
\varphi_{3}(-c) = 0.
\end{array}                                            
\end{equation}

3) In region 1 the particle is located on level $E_{2}$ and this level
is absent in regions 2 and 3. Also in regions 1, 2 and 3 there are
the equal levels $E_{2}^{\prime}$. On this case the particle can not
propagate along level $E_{2}$ to right well and to region 3 of barrier
(there is a full reflection of particle along level $E_{2}$). But the
particle can make transition on level $E_{2}{\prime}$ in region 1 and then
it can propagate to regions 2 and 3 along this level. One can find the
eigenvalue of level $E_{2}$ from the following system:

\begin{equation}
\begin{array}{ll}
\varphi_{1}(-d) = 0, & \varphi_{1}(-c) = 0.
\end{array}
\end{equation}                                       

The eigenvalue of level $E_{2}{\prime}$ satisfies system (2.1)
and system (2.2).

More concretely the first case will be considered in one of the next
sections. Note, that at symmetry of potential ($d=b$, $c=a$,
$W_{0}=0$) for every level of left well one can find the appropriate
level in right well and on the contrary. Therefore, for cases
considered above only the first case is possible for this potential.

Now we consider potential which has rounded off form (see Fig.~\ref{f2}).
Let the particle be localized in the left well. Among possible cases, in
which the particle can propagate through barrier to right well,
we select the following cases:

1) In regions 1 and 2 there are the equal levels $E_{0}$. Then
particle can tunnel from left well to right well along level
$E_{0}$. The eigenvalue for this level can be found from system
(2.1) and the following system:

\begin{equation}
\begin{array}{ll}
\varphi_{1}(0) = \varphi_{2}(0), &
\varphi_{1}^{\prime}(0) = \varphi_{2}^{\prime}(0).
\end{array}
\end{equation}                                       

2) In left well the particle is located on level $E_{1}$ and this
level is absent in region 2. But in regions 1 and 2 there are the
equal levels $E_{1}^{\prime}$. Then particle can not tunnel from
the left well into the right one along level $E_{1}$. But at first
it can make transition from level $E_{1}$ to level $E_{1}^{\prime}$
in region 1 and then it can propagate to region 2 along level
$E_{1}^{\prime}$. System for finding eigenvalue of level $E_{1}$
is given by

\begin{equation}
\begin{array}{ll}
\varphi_{1}(-d) = 0, & \varphi_{1}(0) = 0.
\end{array}
\end{equation}                                       

Eigenvalue for level $E_{1}^{\prime}$ can be obtained from system
(2.1) and system (2.6).

In one of the next sections the problem with potential of such form
will be considered more concretely. As an example, the double-well Morse's
potential is selected. Also note, that for potential of rounded off
form (as in case of the squared potential) only the first case is
possible when the potential is symmetric.

\section {Dependence of distance between two closely located \
levels \ on \ transmission \ coefficient \newline
through barrier in quasi-classic
symmetric potential}
\setcounter{equation}{0}

\begin{figure}[p]
\centering
\includegraphics[width=7cm]{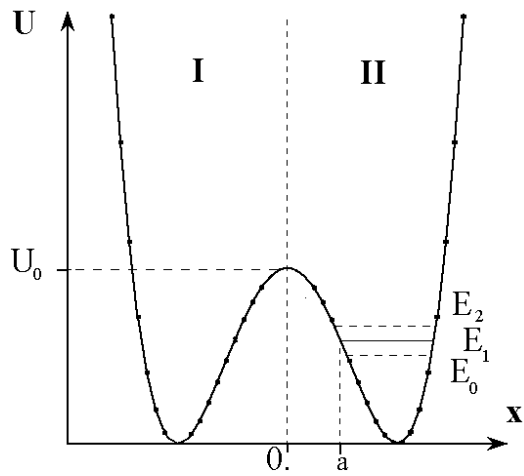}
\caption{Splitting of energy levels in symmetric potential}
\label{f3}
\end{figure}

Consider potential $U(x)$, which has two symmetric wells separated
by barrier (see Fig.~\ref{f3}). If barrier is not penetrable, then there
are energy levels $E_{0}$ corresponding to oscillations of particle
only in one well. The possibility of particle transitions between
wells leads to splitting of every level $E_{0}$ into two closely
located levels $E_{1}$ and $E_{2}$.

We consider the case, when potential $U(x)$ is quasi-classic. Then
the splitting value $\Delta E$ can be obtained through wave function
$\varphi_{0}(x)$, which determined with accuracy to first order
terms in $\hbar$, as follows \cite{Landau.1989}

\begin{equation}
\Delta E = E_{2} - E_{1} = \frac{2\hbar^{2}}{m}
           \varphi_{0}(0)\varphi_{0}^{\prime}(0)
         = \frac{w\hbar}{\pi} * \exp \biggl( -\frac{1}{\hbar}
           \int\limits_{-a}^{a} |p| dx \biggr),
\end{equation}                                         

where

\begin{equation}
\begin{array}{ll}
\varphi_{0}(0) = \sqrt{\displaystyle\frac{w}{2\pi v_{0}}} \exp\biggl(
               - \displaystyle\frac{1}{\hbar} \int\limits_{0}^{a}|p|
                 dx \biggr), &
\varphi_{0}^{\prime}(0) = \displaystyle\frac{mv_{0}}{\hbar} \varphi_{0}(0),
\end{array}
\end{equation}                                         

and $v_{0}=\sqrt{2(U_{0}-E)/m}$, $p$ is impulse of system, $w=2\pi/T$
is frequency of classic periodic oscillations; $a$ is turning point
corresponding to level $E_{0}$ (see Fig.~\ref{f3}).

Transmission coefficient $D$ through barrier in WKB approximation
is determined by \cite{Landau.1989}

\begin{equation}
D = const* \exp \biggl( -\frac{2}{\hbar} \int\limits_{-a}^{a} |p| dx \biggr),
\end{equation}                                      

where proportionality factor $const$ is determined by accuracy of
WKB approximation and is equal to 1 with accuracy of first order
terms in $\hbar$ (see \cite{Landau.1989}). Taking the expressions
(3.1) and (3.3) into consideration, we can write

\begin{equation}
\Delta E = const * \frac{w\hbar}{\pi} \sqrt{D}
\end{equation}                                      

Here, transmission coefficient $D$ is determined by expression (3.3)
for discrete energy spectrum. In accordance with conditions of using WKB
approximation, the expression (3.4) determining the dependence of
level splitting $\Delta E$ on transmission coefficient $D$ is used
only for small values $D$.

Now we consider some cases, for which there are the exact analytical
solutions.

\section{Double-well infinite squared potential}
\setcounter{equation}{0}

Consider system, potential of which consists from two squared wells
separated by squared barrier of finite height (see Fig.~\ref{f1}).
This potential is given by

\begin{equation}
U(x) = \left\{
\begin{array}{rll}
\infty, & \mbox{for } x<-d, x>b; & \\
0,      & \mbox{for } -d<x<-c,   & \mbox{(region I)}; \\
U_{0},  & \mbox{for } -c<x<a,    & \mbox{(region III)}; \\    
-W_{0}, & \mbox{for }  a<x<b,    & \mbox{(region II)}.
\end{array} \right.
\end{equation}

In case of discrete energy spectrum in region $U_{0}>E>0$ we find
the solution of stationary Schr\"{o}dinger equation in form:

\begin{equation}
\varphi(x) = \left\{
\begin{array}{ll}
a_{1} \sin(k(x+d)),                  & \mbox{for } -d<x<-c, \\
a_{2}e^{\chi x} + b_{2}e^{-\chi x},  & \mbox{for } -c<x<a,  \\
a_{3} \sin(k_{3}(x-b)),              & \mbox{for }  a<x<b.
\end{array} \right.
\end{equation}                                             

Here, the following coefficients are used:

\begin{equation}
\begin{array}{l}
k = \frac{1}{\hbar} \sqrt{2mE}, \\
k_{3} = \frac{1}{\hbar} \sqrt{2m(E+W_{0})}, \\     
\chi = \frac{1}{\hbar} \sqrt{2m(U_{0}-E)}.
\end{array}
\end{equation}

Let's consider the case of particle oscillations between wells along
one energy level (without transitions between energy levels). Unknown
coefficients and these energy levels can be found from continuity
conditions of wave function in boundary points $x=-c$, $x=a$ and from
the following normalization condition:

\begin{equation}
\int\limits_{-\infty}^{+\infty}|\varphi(x)|^{2} dx = 1.     
\end{equation}

In result, we obtain unknown coefficients:

\begin{eqnarray}
& a_{1} = \Biggl\{ \displaystyle\frac{d-c}{2} + \displaystyle\frac{a+c}
          {2} * \biggl[ \sin^{2}(k(d-c)) - \displaystyle\frac{k^{2}}
          {\chi^{2}} \cos^{2}(k(d-c)) \biggr] + & \nonumber \\
& + \biggl( \displaystyle\frac{\sin(k(d-c)) - k/\chi \cos(k(d-c))}
    {\sin(k_{3}(b-a)) + k_{3}/\chi \cos(k_{3}(b-a))} \biggr) ^{2} *
    e^{-2\chi(a+c)} \biggl[ \displaystyle\frac{b-a}{2} +
    \displaystyle\frac{1}{4k_{3}} * & \nonumber \\
& * \sin(2k_{3}(a-b)) \biggr] - \displaystyle\frac{1}{4k}
    \sin(2k(d-c)) + \displaystyle\frac{(\sin(k(d-c)) +
    k/\chi \cos(k(d-c)))^{2}}{8\chi} * & \nonumber \\
& * (e^{2\chi(a+c)} - 1) + \displaystyle\frac{(\sin(k(d-c)) -
    k/\chi \cos(k(d-c)))^{2}} {8\chi} * (1 - e^{-2\chi(a+c)})
    \Biggr\} ^{-1/2}; &
\end{eqnarray}                                       

\begin{equation}
\begin{array}{l}
a_{2} = a_{1} * \frac{1}{2} e^{\chi c} (\sin(k(d-c))
      + k/\chi\cos(k(d-c))), \\
b_{2} = a_{1} * \frac{1}{2} e^{-\chi c} (\sin(k(d-c))
      - k/\chi\cos(k(d-c))),  \\
a_{3} = a_{1} * e^{-\chi(a+c)} \displaystyle \frac{k/\chi\cos(k(d-c))
      - \sin(k(d-c))} {k_{3}/\chi\cos(k_{3}(b-a)) + \sin(k_{3}(b-a))}.
\end{array}                                             
\end{equation}

Eigenvalue equation for this potential are given by

\begin{equation}
\begin{array}{l}
E = \hbar^{2} k^{2} / 2m, \\
\displaystyle
\frac{\sin(k_{3}(b-a)) - k_{3}/\chi\cos(k_{3}(b-a))}
     {\sin(k_{3}(b-a)) + k_{3}/\chi\cos(k_{3}(b-a))} *
\displaystyle
\frac{\sin(k(d-c)) - k/\chi\cos(k(d-c))}
     {\sin(k(d-c)) + k/\chi\cos(k(d-c))} = e^{2\chi(a+c)}.
\end{array}                                              
\end{equation}

Now we consider symmetric case $(d=b, c=a, W_{0}=0)$
\cite{Landau.1989,Razavy.1988.PRPLC}.
Wave function became symmetric or antisymmetric:

\begin{equation}
\varphi_{n}(x) = \left\{
\begin{array}{ll}
a_{1} \sin(k_{n}(x+b)),                   & \mbox{for } -b<x<-a; \\
b_{2} ((-1)^{n}e^{\chi x} + e^{-\chi x}), & \mbox{for } -a<x<a; \\
a_{1} (-1)^{n} \sin(k_{n}(b-x)),          & \mbox{for }  a<x<b.
\end{array} \right.                                     
\end{equation}

where $n \in 0,N$. Unknown coefficients $a_{1}$ and $b_{2}$,
obtained from expressions (4.5) and (4.6), are given by

\begin{equation}
\begin{array}{l}
a_{1} = \Biggl\{ b-a + \displaystyle\frac{(\sin(k(b-a)) -
        k/\chi \cos(k(b-a)))^{2}}{4e^{2\chi a}} * \biggl ((-1)^{n}2a +
        \displaystyle\frac{e^{2\chi a} - e^{-2\chi a}}{\chi} \biggr)
        \Biggr\}^{-1/2} \\
b_{2} = a_{1} * \displaystyle\frac{\sin(k(b-a)) -
        k/\chi \cos(k(b-a))}{2e^{\chi a}}.
\end{array}                                                 
\end{equation}

The wave vector $k$ is transformed to form:

\begin{equation}
k = \displaystyle\frac{1}{b-a} * \Biggl\{ -\arcsin\Biggl[
    \displaystyle\frac{\mathstrut 1}{\sqrt{1 +
    \displaystyle\frac{\chi^{2}}{k^{2}} \biggl(
    \displaystyle\frac{(-1)^{n} - e^{2\chi a}}{(-1)^{n} +
    e^{2\chi a}} \biggr) ^{2} }} \Biggr] + \pi n \Biggr\}.
\end{equation}

Consider the particle propagating from the left to right in the
potential (4.1) with asymptotic velocity $\hbar k/m $ and energy
$E<U_{0}$. For it the wave function can be written as follows

\begin{equation}
\varphi(x) = \left\{
\begin{array}{ll}
(\hbar k / m)^{-1/2} * (e^{ikx} - A e^{-ikx}), & \mbox{for }  x<-c; \\
(\hbar k / m)^{-1/2} * (B e^{\chi x} + B^{\prime}e^{-\chi x}),
                                               & \mbox{for } -c<x<a; \\
(\hbar k / m)^{-1/2} * C e^{ikx},              & \mbox{for }   x>a.
\end{array} \right.                                     
\end{equation}

Coefficients $A$, $B$, $B^{\prime}$ and $C$ are obtained from
continuity conditions for $\varphi(x)$ and $d\varphi (x) / d x$ at
points $x = -c$ and $x = a$. The transmission coefficient $D$
and reflection coefficient $R$ calculated as the ratio of the flux
of incident wave in region III to the flux of transmitted
wave in region I or reflected wave in region III, are given by

\begin{equation}
\begin{array}{l}
D = \displaystyle\frac{4kk_{3}\chi^{2}}
    {(kk_{3}+\chi^{2})^{2} \sinh^{2}(\chi(a+c))
    + \chi^{2}(k-k_{3})^{2} ch^{2}(\chi(a+c)) + 4kk_{3}\chi^{2}},  \\
R = \displaystyle\frac{(kk_{3}+\chi^{2})^{2} \sinh^{2}(\chi(a+c))
    + \chi^{2}(k-k_{3})^{2} ch^{2}(\chi(a+c))}
    {(kk_{3}+\chi^{2})^{2} \sinh^{2}(\chi(a+c))
    + \chi^{2}(k-k_{3})^{2} ch^{2}(\chi(a+c)) + 4kk_{3}\chi^{2}}.  \\
\end{array}                                                 
\end{equation}

We find the values $D$ and $R$ for transmission of particle through
barrier in continuous energy spectrum. But comparison of transmission
coefficient $D$ with its small values determined by expression (4.11)
for symmetric case
with transmission coefficient $D$ determined by expression (3.3), which
is obtained in WKB approximation for discrete energy spectrum, show,
that both approaches give identical formulations with accuracy to
normalized constant:
$D=const*\exp(-2\chi a)$, where $const$ is determined by accuracy
of WKB approximation and is equal to 1 in terms of first order in
$\hbar$ \cite{Landau.1989,Anderson.1989.AJPIA}. In this sense, we
will formally
consider the expression (4.11) as the determination of transmission
and reflection coefficients for discrete energy spectrum. The values
$k$ and $\chi$, used in expression (4.11), can be obtained from
equation (4.7) or (4.10).

We analyze the periodicity of particle oscillations between wells in
symmetric potential. For this we consider equation (4.10) which can
be written as

\begin{equation}
\begin{array}{l}
f_{1}(k_{n}) = k_{n}(b-a),  \\
f_{2}(k_{n}) = - \arcsin \Biggl\{ \displaystyle\frac{1}{\sqrt{1 +
               \displaystyle\frac{\chi^{2}}{k^{2}}
               \biggl( \displaystyle\frac{(-1)^{n} - e^{2\chi a}}
               {(-1)^{n} + e^{2\chi a}} \biggr) ^{2} }} \Biggr\} + \pi n.
\end{array}                                         
\end{equation}

\begin{figure}[p]
\centering
\includegraphics[width=7cm]{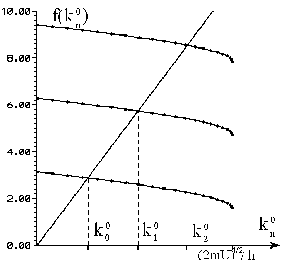}
\includegraphics[width=7cm]{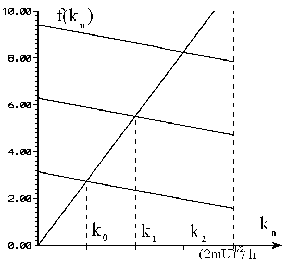}
\caption{The values $k_{n}^{0}$ are exact graphic solutions of
system (4.15) and the values $k_{n}$ are graphic solutions on
this system, where the second equation is changed to its linear
approximation $f_{2} (k_{n})$}
\label{f4}
\end{figure}

The graphic analysis of exact solutions of system (4.13) gives
a number of values $k_{n}^{0}$ (see Fig.~\ref{f4}). Changing the
second equation of system (4.13) to its linear relation $f_{2}
(k_{n})$, which can be obtained by linear approximation, we find
the following values of $k_{n}$:

\begin{equation}
\begin{array}{l}
k_{n} = k_{0} (2n+1),  \\
E_{n} = E_{0} (2n+1)^{2} = E_{0} + 4E_{0} * n(n+1) =
        E_{0} + \Delta * l_{n},  \\
\Delta = 4 * E_{0}.
\end{array}                                        
\end{equation}

where $n,l_{n} \in 0,N$. From expressions (4.14) we obtain period
$T$ of particle oscillations between wells:

\begin{equation}
T = \displaystyle\frac{2\pi\hbar}{\Delta} =
    \displaystyle\frac{\pi\hbar}{2E_{0}} =
    \displaystyle\frac{\pi\hbar}{2} * \frac{(2n+1)^{2}}{E_{n}} =
    \displaystyle\frac{\pi m}{\hbar k_{n}^{2}} * (2n+1)^{2}.
\end{equation}                                  

For values $k_{n}$ the accuracy $\pi/(2(b-a))$ is used. The series
of solutions $k_{n}$ is bounded by maximum value $k_{N}$, where $N$
(the number of energy levels in region $E<U_{0}$) satisfies the
following condition:

\[ 0 < N < \displaystyle\frac{\sqrt{2mU_{0}} - \hbar k_{0}}{2\hbar k_{0}}.\]

From expression (4.15) we obtain:

\begin{equation}
k_{n}^{2} = (2n+1)^{2}\displaystyle\frac{\pi m}{\hbar T} = \Delta *
            \displaystyle\frac{m(2n+1)^{2}}{2\hbar^{2}}.
\end{equation}                                          

Using expression (4.12) for symmetric case and expression (4.16),
we find the dependence of transmission coefficient $D$ through
barrier on oscillation period $T$ and on value $\Delta$:

\begin{equation}
D_{n}(T) = \Biggl\{ 1 + \displaystyle\frac{\sinh^{2}(2a \sqrt{
           \displaystyle\frac{2mU_{0}}{\hbar^{2}} -
           (2n+1)^{2} \displaystyle\frac{\pi m}{\hbar T}})}
           {4 (2n+1)^{2} \displaystyle\frac{\pi m}{\hbar T} \biggl(
           1 - (2n+1)^{2} \frac{\pi\hbar}{2U_{0}T} \biggr)} \Biggr\} ^{-1}.
\end{equation}                                            

\begin{equation}
D_{n}(\Delta) = \Biggl\{ 1 + \displaystyle\frac{\sinh^{2}(2a \sqrt{
                \displaystyle\frac{2mU_{0}}{\hbar^{2}} -
                \Delta * \displaystyle\frac{(2n+1)^{2}}{2\hbar^{2}}} )}
                {\Delta * \displaystyle\frac{2m(2n+1)^{2}}{\hbar^{2}}
                \biggl(1 - \Delta * \displaystyle\frac{(2n+1)^{2}}{4U_{0}}
                \biggr)} \Biggr\}^{-1}.
\end{equation}                                            

Also note, that in both limiting cases $D \to 0$ and $D \to \infty$ the
oscillation period $T$ approaches a value:

\begin{equation}
  T = \displaystyle\frac{4m (b-a)^{2}}{\pi\hbar}
\end{equation}                                            

\section{Double-well Morse's potential}
\setcounter{equation}{0}

Consider the system, potential of which is shown in Fig.~\ref{f2}
and is given by

\begin{equation}
U(x) = \left\{
\begin{array}{ll}
A(e^{-2\alpha(x+c)} - 2e^{-\alpha(x+c)}), & \mbox{for } -d<x<0, \\
B(e^{2\beta(x-a)} - 2e^{\beta(x-a)}),     & \mbox{for  }  0<x<b.
\end{array} \right.
\end{equation}                                         

For this potential we find solutions of stationary Schr\"{o}dinger
equation. Performing the following changes:

\begin{equation}
\begin{array}{rl}
\xi_{A} = \displaystyle\frac{2\sqrt{2mA}}{\alpha\hbar} e^{-\alpha(x+c)},
        & \mbox{for } -d<x<0, \\
\xi_{B} = \displaystyle\frac{2\sqrt{2mB}}{\beta\hbar} e^{\beta(x-a)},
        & \mbox{for  }  0<x<b,
\end{array}
\end{equation}                                         

and introducing the following notations:

\begin{equation}
\begin{array}{ll}
s_{A} = \displaystyle\frac{\sqrt{-2mE}}{\alpha\hbar}, &
n_{A} = \displaystyle\frac{\sqrt{2mA}}{\alpha\hbar} -
        \biggl ( s_{A} + \displaystyle\frac{1}{2} \biggr ), \\
s_{B} = \displaystyle\frac{\sqrt{-2mE}}{\beta\hbar}, &
n_{B} = \displaystyle\frac{\sqrt{2mB}}{\beta\hbar} -
        \biggl ( s_{B} + \displaystyle\frac{1}{2} \biggr ),
\end{array}                                            
\end{equation}

we transform the stationary Schr\"{o}dinger equation to the following
form:

\begin{equation}
  \biggl\{ \displaystyle\frac{d^{2} \varphi}{d\xi^{2}} +
  \displaystyle\frac{1}{\xi}\displaystyle\frac{d\varphi}{d\xi} +
  \biggl(-\displaystyle\frac{1}{4} + \displaystyle\frac{n + s + 1/2}
  {\xi} - \displaystyle\frac{s^{2}}{\xi^{2}} \biggr)
  \varphi\biggr\}\biggr|_{A,B} = 0.
\end{equation}                                         

Here, the indexes $A$ or $B$ are used by values $\varphi$, $\xi$,
$n$ and $s$ with dependence on that, the equation (5.4) is solved
in region of left or right well, respectively. For obtaining the
general solution, let's omit these indexes. Using the following
changes:

\begin{equation}
\begin{array}{lcl}
y(\xi) & = & \xi^{-\frac{c}{2} + \frac{1}{2}} e^{\frac{\xi}{2}}
             \varphi(\xi), \\
c      & = & 1 + 2s,
\end{array}
\end{equation}                                         

the equation (5.4) can be transformed to confluent hypergeometric
equation as follows

\begin{equation}
\xi \displaystyle\frac{d^{2}y}{d\xi^{2}} +
(c-\xi) \displaystyle\frac{dy}{d\xi} + ny = 0.
\end{equation}                                         

The particular solutions of this equation can be written by use of
confluent hypergeometric function $F(-n,c; \xi)$
\cite{Beitman.1973,Abramowitz.1979}:

\begin{equation}
\begin{array}{rcl}
y_{1}(\xi) & = & F(-n,c;\xi), \\
y_{2}(\xi) & = & \xi^{1-c} F(-n-c+1,2-c;\xi), \\
y_{3}(\xi) & = & e^{\xi} F(c+n,c;-\xi), \\
y_{4}(\xi) & = & \xi^{1-c} e^{\xi} F(1+n,2-c;-\xi).
\end{array}
\end{equation}                                         

Consider the case $c \not\in Z$ (i.e. $2s \not\in Z$). Then both
the particular solutions $y_{3}$ and $y_{4}$ are linear dependent
on $y_{1}$ and $y_{2}$, because they are transformed to $y_{1}$
and $y_{2}$ by Kummer's transformation
\cite{Beitman.1973,Abramowitz.1979}:

\begin{equation}
F(a,c; \xi) = e^{\xi} F(c-a,c; -\xi).
\end{equation}                                         

Solutions $y_{1}$ and $y_{2}$ are linear independent between themselves.
It can see this from its behavior close by $\xi \to 0$. Then the general
solution of equation (5.6) can be written as follows

\begin{equation}
y(\xi) = c_{1} F(-n,c; \xi) + c_{2} \xi^{1-c} F(-n-c+1,2-c; \xi)
\end{equation}                                         

where $c_{1}$ and $c_{2}$ are arbitrary constants. In initial variables
the general solution has form:

\begin{equation}
\varphi(\xi) = e^{-\frac{\xi}{2}} [c_{1} \xi^{s} F(-n,1+2s; \xi) +
               c_{2} \xi^{-s} F(-n-2s,1-2s; \xi)]
\end{equation}                                         

Expression (5.10) with relations (5.2) and (5.3) is general solution
of stationary Schr\"{o}dinger equation at $2s \not\in Z$ and $E<0$ for
regions of left and right wells. Coefficients $c_{1A}$, $c_{1B}$,
$c_{2A}$ and $c_{2B}$ (where indexes $A$ or $B$ for coefficients
$c_{1}$ and $c_{2}$ indicate at left or right well, respectively)
are obtained from boundary conditions and normalization condition.
Imposed boundary conditions determine the energy spectrum as a
discrete ones and using them it can find the energy eigenvalues
of system.

Discrete energy spectrum determines the requirement that wave
function will be finited all over the region of its determination.
Finity of wave function in regions $-d<x<0$ and
$0<x<b$ depends on constraining condition of particular solutions,
by use of which the general solution (5.10) can be represented.
Constraining of particular solutions at finite values $x$ depends on
constraining of functions $F(-n,1+2s;\xi)$ and $F(-n-2s,1-2s;\xi)$. But
these functions are confluent hypergeometric and are represented by
converging series at finite $\xi$ \cite{Abramowitz.1979,Vorobiev.1986}.
At $n = 0$ or $n \in N$ the first function has form of polynomial, and
therefore, it is bounded (because values $\xi$ are finited). In this
fashion, at $n+2s=0$ or $n+2s \in N$ the second function is bounded.
Therefore, both the particular solutions are bounded all over the
region $-d<x<b$ (factor $\xi^{-s}$ at $\xi \to 0$ is finited because
of constraining condition at $x=0$). Therefore, the general solution
(5.10) is also bounded all over the region $-\infty<x<\infty$.

Since wave function satisfies condition of constraining all
over the region of definition $x$, it can be normalized by

\begin{equation}
\int \limits_{-d}^{b} |\varphi(x)|^{2} dx = 1.
\end{equation}                                         

Now we find the energy eigenvalues of system. Under consideration of
possible solutions we select two cases:

1) The particle oscillate between wells along one energy level
(without transitions between energy levels). On this case the energy
levels, which are obtained from eigenvalue equation in region of left
well, must correspond to energy levels, which are obtained from
eigenvalue equation in region of right well. On this case the
following boundary conditions are required:

\begin{equation}
\begin{array}{l}
\varphi_{A}(0) = \varphi_{B}(0), \\
\varphi_{A}^{\prime}(0) = \varphi_{B}^{\prime}(0), \\
\varphi_{A}(-d) = 0, \\
\varphi_{B}(b) = 0,
\end{array}
\end{equation}                                         

where $\varphi_{A}(x)$ and $\varphi_{B}(x)$ are the general solutions
(5.10) of stationary Schr\"{o}dinger equation in regions of left
and right wells, respectively. Solution of equation system (5.12)
gives the eigenvalue equation of form:

\begin{eqnarray}
& - \displaystyle\frac{\alpha \xi_{A0}}{\beta \xi_{B0}} \biggl[
    \displaystyle\frac{c_{1B}}{c_{2B}} * \xi_{B0}^{s_{B}}
    F(-n_{B}, 1+2s_{B}; \xi_{B0}) + \xi_{B0}^{-s_{B}}
    F(-n_{B}-2s_{B}, 1-2s_{B}; \xi_{B0}) \biggr] *
  & \nonumber \\
& * \biggl \{ \displaystyle\frac{c_{1A}}{c_{2A}} * \biggl [ -
    \displaystyle\frac{1}{2} \xi_{A0}^{s_{A}}
    F(-n_{A}, 1+2s_{A}; \xi_{A0}) + s_{A} * \xi_{A0}^{s_{A}-1}
    F(-n_{A}, 1+2s_{A}; \xi_{A0}) + & \nonumber \\
& + \xi_{A0}^{s_{A}} \displaystyle\frac{\partial
    F(-n_{A}, 1+2s_{A}; \xi_{A0})} {\partial \xi_{A}} \biggr] +
    \biggl [ - \displaystyle\frac{1}{2} \xi_{A0}^{-s_{A}}
    F(-n_{A}-2s_{A}, 1-2s_{A}; \xi_{A0}) - & \nonumber \\
& - s_{A} * \xi_{A0}^{-s_{A}-1} F(-n_{A}-2s_{A}, 1-2s_{A}; \xi_{A0}) +
    \xi_{A0}^{-s_{A}} \displaystyle\frac{ \partial
    F(-n_{A}-2s_{A}, 1-2s_{A}; \xi_{A0})}
    {\partial \xi_{A}} \biggr] \biggr\} = & \nonumber \\
& = \biggl \{ \displaystyle\frac{c_{1B}}{c_{2B}} *
    \biggl [ - \displaystyle\frac{1}{2} \xi_{B0}^{s_{B}}
    F(-n_{B}, 1+2s_{B}; \xi_{B0}) + s_{B} * \xi_{B0}^{s_{B}-1}
    F(-n_{B}, 1+2s_{B}; \xi_{B0}) + & \nonumber \\
& + \xi_{B0}^{s_{B}} \displaystyle\frac{ \partial
    F(-n_{B}, 1+2s_{B}; \xi_{B0})} {\partial \xi_{B}} \biggr ] +
    \biggl [ - \displaystyle\frac{1}{2} \xi_{B0}^{-s_{B}}
    F(-n_{B}-2s_{B}, 1-2s_{B}; \xi_{B0}) - & \nonumber \\
& - s_{B} * \xi_{B0}^{-s_{B}-1} F(-n_{B}-2s_{B}, 1-2s_{B}; \xi_{B0}) +
    \xi_{B0}^{-s_{B}} \displaystyle\frac{ \partial
    F(-n_{B}-2s_{B}, 1-2s_{B}; \xi_{B0})}
    {\partial \xi_{B}} \biggr ] \biggr \} * & \nonumber \\
& * \biggl [ \displaystyle\frac{c_{1A}}{c_{2A}} * \xi_{A0}^{s_{A}}
    F(-n_{A}, 1+2s_{A}; \xi_{A0}) + \xi_{A0}^{-s_{A}}
    F(-n_{A}-2s_{A}, 1-2s_{A}; \xi_{A0}) \biggr ] &
\end{eqnarray}                                         

where

\begin{equation}
\begin{array}{l}
  \displaystyle\frac{c_{1A}}{c_{2A}} = - \Bigl ( \xi_{A0}e^{\alpha d}
  \Bigr )^{-2s_{A}} * \displaystyle\frac{
  F(-n_{A}-2s_{A}, 1-2s_{A}; \xi_{A0}e^{\alpha d})}
  {F(-n_{A}, 1+2s_{A}; \xi_{A0}e^{\alpha d})}, \\
  \displaystyle\frac{c_{1B}}{c_{2B}} = - \Bigl ( \xi_{B0}e^{\beta b}
  \Bigr )^{-2s_{B}} * \displaystyle\frac{
  F(-n_{B}-2s_{B}, 1-2s_{B}; \xi_{B0}e^{\beta b})}
  {F(-n_{B}, 1+2s_{B}; \xi_{B0}e^{\beta b})}, \\
\end{array}
\end{equation}                                         

\begin{equation}
\begin{array}{l}
\xi_{A0} = \displaystyle\frac{2\sqrt{2mA}}{\alpha \hbar} * e^{-\alpha c}, \\
\xi_{B0} = \displaystyle\frac{2\sqrt{2mB}}{\beta \hbar} * e^{-\beta a}.
\end{array}
\end{equation}                                         

Transform expressions (5.3) to form:

\begin{equation}
\begin{array}{l}
s_{B} = \displaystyle\frac{\alpha}{\beta} * s_{A}, \\
n_{B} = \displaystyle\frac{\alpha}{\beta} \sqrt{\displaystyle\frac{B}{A}}
        \Bigl ( n_{A} + s_{A} + \frac{1}{2} \Bigr ) -
        \Bigl ( \displaystyle\frac{\alpha} {\beta} s_{A} +
        \displaystyle\frac{1}{2} \Bigr ).
\end{array}
\end{equation}                                         

To find the eigenvalues, we need to substitute
the expressions (5.14) and (5.16) into equation (5.13) and to
resolve it relatively the value $s_{A}$, which unequivocally
defines eigenvalue $E$. The solution of equation (5.13) is
performed by use of numerical methods and determines the energy
eigenvalues corresponding to particle oscillations between wells.

2) Now consider another case: the particle oscillates in one well
(for example, in left one). On this case its full reflection take
place from the middle of barrier (here, the transition of particle
to another energy level is possible, which exist in both regions,
with further tunneling of particle along it). Note, that on this
case the reflection of particle from barrier is principally possible
along energy levels of range $0>E>U_{0}$ (this case of reflection
is impossible for symmetric potential). We determine the condition
of particle reflection from barrier as follows

\begin{equation}
\varphi_{A}(0) = 0.
\end{equation}                                         

Using this condition and also the following boundary condition

\begin{equation}
\varphi_{A}(-d) = 0.
\end{equation}                                         

we obtain equation, from which it can find the energy eigenvalues for
particle oscillations in left well:

\begin{eqnarray}
  e^{2s_{A} \alpha d} & * F(-n_{A}-2s_{A}, 1-2s_{A}; \xi_{A0}) *
F(-n_{A}, 1+2s_{A}; \xi_{A0} e^{\alpha d}) = &  \nonumber \\
= & F(-n_{A}-2s_{A}, 1-2s_{A}; \xi_{A0} e^{\alpha d}) *
F(-n_{A}, 1+2s_{A}; \xi_{A0})  &
\end{eqnarray}                                         

Resolving this equation relatively unknown values $s_{A}$, it can
find the energy eigenvalues of system. In this fashion, it can obtain
equation which determines the energy eigenvalues for particle
oscillations in right well:

\begin{eqnarray}
  e^{2s_{B} \beta b} & * F(-n_{B}-2s_{B}, 1-2s_{B}; \xi_{B0}) *
F(-n_{B}, 1+2s_{B}; \xi_{B0} e^{\beta b}) = &  \nonumber \\
= & F(-n_{B}-2s_{B}, 1-2s_{B}; \xi_{B0} e^{\beta b}) *
F(-n_{B}, 1+2s_{B}; \xi_{B0})  &
\end{eqnarray}                                         

Equations (5.19) and (5.20) include the case when some energy levels
of left well are equal to some energy levels of right well. On this
case the following condition is fulfilled:

\[
\varphi_{A}(0) = \varphi_{B}(0) = 0, \\
\]

\[
\varphi_{A}^{\prime}(0) = \varphi_{B}^{\prime}(0) = 0,
\]

which correspond to particle oscillations between wells. Therefore,
it need to except such energy levels from analysis of particle
behavior in one well.

Consider symmetric potential. On this case every energy level obtained
by solution of eigenvalue equation in region of left well, is equal
to corresponding energy level which is obtained by solution of
eigenvalue equation in region of right well. Therefore, the particle
located on any energy level of region $E<0$ will be oscillated between
wells. Wave function becomes symmetric or antisymmetric. System of
equations for finding energy eigenvalues is given by

\[ \varphi(-d) = 0, \varphi(0) = 0, \]

if wave function is symmetric (even states), and

\[ \varphi(-d) = 0, \varphi^{\prime}(0) = 0, \]

if wave function is antisymmetric (odd states).

\section{Symmetric potential of form $x^{2}+B^{2}/x^{2}$}
\setcounter{equation}{0}

\begin{figure}[p]
\centering
\includegraphics[width=7cm]{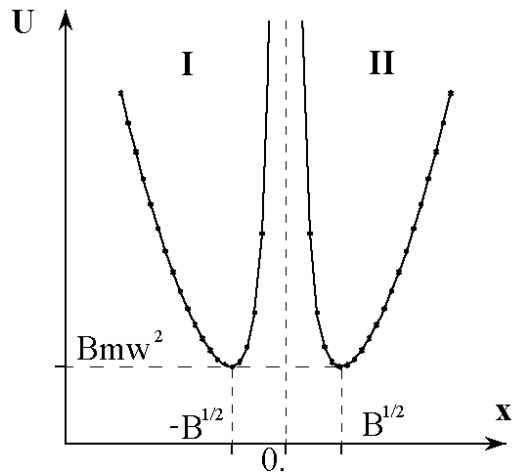}
\caption{Symmetric potential of form $x^{2} + B^{2}/x^{2}$}
\label{f5}
\end{figure}

To analyze the particle behavior in symmetric potential with enough
high barrier, it can use potential of the following form:

\begin{equation}
U(x) = \frac{mw^{2}}{2} \biggl( x^{2} + \frac{B^{2}}{x^{2}} \biggr)
\end{equation}                                         

where $x \in ]-\infty;+\infty[$. This potential is shown on
Fig.~\ref{f5}. Use new parameters:

\begin{equation}
\begin{array}{rcl}
G & = & \displaystyle\frac{2mE}{\hbar^{2}}, \\
F & = & -\displaystyle\frac{m^{2}w^{2}}{\hbar^{2}}, \\
K & = & -\displaystyle\frac{m^{2}w^{2}}{\hbar^{2}} * B^{2}.
\end{array}
\end{equation}                                         

Then stationary Schr\"{o}dinger equation is transformed to form:

\begin{equation}
  \frac{d^{2} \varphi}{dx^{2}} + \biggl(G + Fx^{2} +
  \frac{K}{x^{2}}\biggr) * \varphi = 0
\end{equation}                                         

Let's find solutions of this equation. Perform the following change
of variables:

\begin{equation}
\begin{array}{l}
\xi = \alpha x^{2}, \\
\alpha = \displaystyle\frac{mw}{\hbar} = \sqrt{-F}, \\
\varphi(\xi) = (\xi / \alpha)^{-1/4} w(\xi).
\end{array}
\end{equation}                                         

In result, the equation (6.3) is transformed to standard Whittaker's
form \cite{Beitman.1973,Abramowitz.1979}:

\begin{equation}
  \displaystyle\frac{d^{2}w}{d\xi^{2}} + \biggl[ -\displaystyle\frac{1}{4} +
  \displaystyle\frac{G}{4\sqrt{-F}} \displaystyle\frac{1}{\xi} +
  \biggl( \displaystyle\frac{3}{16} + \displaystyle\frac{K}{4} \biggr)
  \displaystyle\frac{1}{\xi^{2}} \biggr] w = 0
\end{equation}                                         

Using the following parameters and performing the following change
of variables:

\begin{equation}
\begin{array}{lcl}
   k    & = & \displaystyle\frac{G}{4 \sqrt{-F}}, \\
\mu^{2} & = & \displaystyle\frac{1}{16} - \displaystyle\frac{K}{4}, \\
   a    & = & \displaystyle\frac{1}{2} - k + \mu, \\
   c    & = & 1 + 2 \mu,
\end{array}
\end{equation}                                         

\begin{equation}
 y(\xi) = \xi^{-c/2}e^{\xi/2}w(\xi),
\end{equation}                                         

we transform the equation (6.5) to confluent hypergeometric
equation of form:

\begin{equation}
  \xi \displaystyle\frac{d^{2}y}{d\xi^{2}} + (c - \xi)
  \displaystyle\frac{dy}{d\xi} - ay = 0
\end{equation}                                         

Particular solutions of this equation can be represented by confluent
hypergeometric function $F(a,c;\xi)$ as follows

\begin{equation}
\begin{array}{rcl}
y_{1}(\xi) & = & F(a,c;\xi), \\
y_{2}(\xi) & = & \xi^{1-c} F(a-c+1,2-c;\xi), \\
y_{3}(\xi) & = & e^{\xi} F(c-a,c;-\xi), \\
y_{4}(\xi) & = & \xi^{1-c} e^{\xi} F(1-a,2-c;-\xi).
\end{array}
\end{equation}                                         

Let's consider the case $c \not\in Z$. In accordance with Kummer's
transformation \cite{Beitman.1973,Abramowitz.1979}, the solutions $y_{3}$
and $y_{4}$ can be written through $y_{1}$ and $y_{2}$. Therefore,
the solutions $y_{3}$ and $y_{4}$ are linear depended on $y_{1}$ and
$y_{2}$. Write first two solutions $y_{1}$ and $y_{2}$ in initial
variables:

\begin{equation}
\begin{array}{l}
\varphi_{1}(x) = \alpha^{1/2 + \mu} x^{-1+2\mu} e^{-\alpha x^{2} / 2}
                 F\left(a, 1+2\mu; \alpha x^{2}\right), \\
\varphi_{2}(x) = \alpha^{1/2 + \mu} x^{-1} e^{-\alpha x^{2} / 2}
                 F\left( a-2\mu, 1-2\mu; \alpha x^{2} \right).
\end{array}
\end{equation}                                         

Condition of wave function finity all over the range $x$ requires,
that the following conditions are fulfilled:

\begin{equation}
\begin{array}{l}
\mbox{for } \varphi_{1}(x) : a \in 0, - N; 2 \mu \not\in - N, \\
\mbox{for } \varphi_{2}(x) : -a+2\mu \in 0,N; 2 \mu \not\in N.
\end{array}
\end{equation}                                         

As a result of these conditions, the series, which represent confluent
hypergeometric function for solutions $\varphi_{1}$ and $\varphi_{2}$,
transform to polynomial and the spectrum becomes discrete.
Analysis of expressions (6.11) shows, that solutions $\varphi_{1}$
and $\varphi_{2}$ can not be used at same time. But both solutions
have equal energy eigenvalues $E$ which can be written as

\begin{equation}
E_{n}^{\pm} = 2\hbar w \biggl(\displaystyle\frac{1}{2} + n \pm
              \displaystyle\frac{1}{4} \sqrt{1 +
              \displaystyle\frac{4m^{2}w^{2}}{\hbar^{2}} B^{2}}\biggr)
\end{equation}                                         

where $n \in 0,N$. The general solution of wave function can be
represented through $\varphi_{1}$ or $\varphi_{2}$ (which are linear
independed between themselves). Both energy eigenvalues $E_{n}^{+}$
and $E_{n}^{-}$ used in expression (6.12) can be considered separately
as defining two independed waves with oscillation periods $T^{+}$
and $T^{-}$, respectively. To obtain period for every wave, we write
the equation (6.12) as follows

\[
E_{n}^{\pm} = E_{0}^{\pm} + \Delta * n = 2 \hbar w * n + 2\hbar w *
  \biggl( \displaystyle\frac{1}{2} \pm \displaystyle\frac{1}{4}
  \sqrt{1 + \displaystyle\frac{4m^{2}w^{2}}{\hbar^{}2} B^{2}} \biggr)
\]

From this we obtain

\begin{equation}
\begin{array}{lcl}
E_{0}^{\pm} & = & 2\hbar w * \biggl( \displaystyle\frac{1}{2} \pm
                  \displaystyle\frac{1}{4} \sqrt{1 +
                  \displaystyle\frac{4m^{2}w^{2}}{\hbar^{}2}
                  B^{2}} \biggr), \\
\Delta      & = & 2\hbar w, \\
T^{\pm}     & = & \displaystyle\frac{2 \pi \hbar}{\Delta} =
                  \displaystyle\frac{\pi}{w}.
\end{array}
\end{equation}                                         

It can see from expressions (6.13), that two considered waves have
equal periods. In accordance with the following relation

\[
A e^{iwt + i\delta_{1}} + B e^{iwt + i\delta{2}} = e^{iwt}
  \bigl( A e^{i\delta_{1}} + B e^{i\delta_{2}} \bigr),
\]

it can see that general period of oscillations between wells is equal
to $T^{+}$ or $T^{-}$:

\begin{equation}
T = T^{\pm} = \displaystyle\frac{\pi}{w}.
\end{equation}                                         

Now we find dependence of transmission coefficient $D$ through
barrier on oscillation period $T$ between wells. Consider the case
of small values $D$, when expression (3.4) can be used. From
expression (6.13) we find the distance between two closely located
levels by

\begin{equation}
\Delta E = E_{n}^{+} - E_{n}^{-} = \hbar w \sqrt{1 +
           \displaystyle\frac{4m^{2}w{2}}{\hbar^{2}} B^{2}}
\end{equation}                                         

Using this relation and also expression (3.4), we obtain the
transmission coefficient as follows

\begin{eqnarray}
D & \sim & \pi^{2} \biggl( 1 + \displaystyle\frac{4m^{2} w^{2}}
  {\hbar^{2}} B^{2} \biggr) =
  \pi^{2} \biggl( 1 + \displaystyle\frac{m^{2} B^{2}} {\hbar^{4}}
  \Delta^{2} \biggr) = \nonumber \\
  &  =   & \pi^{2} \biggl( 1 + \displaystyle\frac{\pi^{2}16 m^{2}
  B^{2}}{\hbar^{2}} \displaystyle\frac{1}{T^{2}} \biggr).
\end{eqnarray}                                         

This expression establishes the dependencies of transmission coefficient
$D$ on oscillation period $T$ between wells and on largest common
divizor $\Delta$ which is determined by expression (1.1).

\section{Double-well symmetric parabolic potential}
\setcounter{equation}{0}

Consider the system, potential of which can be written as

\begin{equation}
U(x) = \left\{
\begin{array}{rl}
\frac{1}{2}mw^{2}(x+a)^{2}, & \mbox{for } x<0; \\
\frac{1}{2}mw^{2}(x-a)^{2}, & \mbox{for } x>0.
\end{array} \right.
\end{equation}                                         

Let's assume, that potential $U(x)$ satisfies conditions of using
WKB approximation. In result of tunneling through barrier the
displacements of energy levels, which takes place because of level
splitting, from their positions without tunneling are determined by
expression (3.1). For potential (7.1) the energy eigenvalues are
given by

\begin{equation}
\begin{array}{l}
  E_{n}^{-} = \hbar w (n+\frac{1}{2}) - \Delta E_{n}, \\
  E_{n}^{+} = \hbar w (n+\frac{1}{2}) + \Delta E_{n},  \\
  \Delta E_{n} = |E_{n}^{\pm} - E_{n}^{(0)}| =
    \displaystyle\frac{\hbar^{2}}{m} \varphi_{n}^{0}(0)
    \displaystyle\frac{\partial\varphi_{n}^{(0)}(0)}{\partial x},
\end{array}
\end{equation}                                         

where $E_{n}^{-}$ and $E_{n}^{+}$ are eigenvalues,
which occur because of splitting and correspond to symmetric and
antisymmetric wave function, respectively. $\varphi_{n}^{(0)}$
is solution of stationary Schr\"{o}dinger equation in region $x>0$
without splitting and has form:

\begin{equation}
  \varphi_{n}^{(0)}(x) = A_{n} e^{-\alpha^{2}(x-a)^{2} / 2}
                         H_{n}(\alpha(x-a))
\end{equation}                                         

where $\alpha = \sqrt{mw/\hbar}$, $H_{n}(\xi)$ is Hermitian polynomial.
Using (7.3), we find the displacement value:

\begin{equation}
  \Delta E_{n} = \displaystyle\frac{\hbar^{2}}{m} A_{n}^{2}
                 e^{-\alpha^{2} a^{2}} H_{n}(\alpha a)
                 (\alpha^{2}a H_{n}(\alpha a) - 2n H_{n-1}(\alpha a))
\end{equation}                                         

Coefficient $A_{n}$ can be obtained from normalization condition of
$\varphi_{n}^{(0)}(x)$. Function $\varphi_{n}^{(0)}(x)$ is assumed
as normalized one, that integral of $|\varphi_{n}^{(0)}|^{2}$ in region
of right well is equal to 1. Then it can write

\begin{eqnarray}
A_{n}^{2} & = & \biggl\{ \int\limits_{0}^{+\infty} e^{-\alpha^{2}
                (x-a)^{2}} H_{n}^{2}(\alpha(x-a)) dx \biggr\}^{-1}
            =   \nonumber \\
          & = & \displaystyle\frac{\alpha}{2^{n} n!} \biggl\{
                \displaystyle\frac{\sqrt{\pi}}{2} (1 + erf(\alpha a)) -
                e^{-\alpha^{2} a^{2}} \sum_{k=0}^{n-1}
                \displaystyle\frac{H_{n-k}(\alpha a) *
                H_{n-k-1}(\alpha a)}{(n-k)!} \biggr\}^{-1}
\end{eqnarray}                                         

where $erf(x) = \displaystyle\frac{2}{\sqrt{\pi}} \int\limits_{0}^{x}
e^{-\xi^{2}} d\xi$ is integral of probability \cite{Abramowitz.1979}.




\section{Conclusions}

In all problems considered above the attempt to describe the tunneling
process of particle through barrier with its oscillations between wells
in double-well symmetric potential was made on the basis of such
main parameters as period $T$ of particle oscillations between wells,
transmission coefficient $D$ through barrier, reflection coefficient
$R$ from barrier (for squared potential). The transmission coefficient
$D$ through barrier is found with consideration of particle, which is
incident upon the barrier having asymptotic velocity $\hbar k/m$
and is initially determined for continuous energy spectrum (see
\cite{Landau.1989,Razavy.1988.PRPLC}). One can obtain all considered above
parameters and describe the tunneling of particle through barrier from
found energy eigenvalues of considered systems. If the transmission
coefficient is small, then it can be calculated in WKB approximation
for discrete energy spectrum. Comparison of both methods realized for
squared potential shows that the values of transmission coefficient,
calculated by these methods, are equal with accuracy to normalization
constant which is determined by accuracy of used WKB approximation and
equal to 1 for first order terms in $\hbar$. Therefore, the transmission
coefficient of particle through barrier between two wells in double-well
potential is considered only formally, if it was initially determined
for continuous spectrum, or with accuracy to normalization constant,
if it is obtained in WKB approximation.

The dependencies of transmission coefficient when its value is small
on another parameters are identical for both methods. These dependencies
are obtained, if these parameters are the oscillation period between
wells and largest common divizor determined by expression (1.1) when
WKB approximation is applied.

All parameters considered above and describing the tunneling of
particle through barrier can be obtained on the basis of found
energy levels of considered systems. For asymmetric forms of
potential they are more difficult to found.

For asymmetric forms of potentials the splitting of energy levels,
which occurs because of tunneling of particle through barrier
and is studied by theory of WKB approximation, gives some features,
which are absent when potential is symmetric (for example, the
possibility of particle reflection from barrier along level, which
is located higher than height of barrier). In case of potential
asymmetry the use of boundary conditions must be careful enough.
Therefore, in present work a considerable attention is paid to the
problem of finding the solutions of energy levels for asymmetric
potentials.

The problem of squared potential was studied early in literature
(for example, see \cite{Landau.1989,Shiff.1957,
Razavy.1988.PRPLC,Anderson.1989.AJPIA,Senn.1992.AJPIA,Heiss.1989.PHRVA}).
This problem is considered here as
one of the simplest cases because it can give visual teaching picture
for studding the tunneling particle behavior. Relative easiness of
finding the solutions in comparison with problems with rounded off
potentials allows to study deeply the process of tunneling on the basis
of oscillation period and transmission and reflection coefficients.

It is significantly difficult to obtain solutions for potential,
which has rounded off form. Using double-well Morse's potential (as
an example), the analysis of solution existence is studied and
energy eigenvalues equation is obtained, the further solution of
which can be calculated by use of numerical methods.

The interesting symmetric double-well potential of form $x^{2} +
B^{2}/x^{2}$ is also analyzed because the exact analytical solutions
exist for it (and one can obtain energy eigenvalues $E_n$, period
$T$ of oscillations between wells in the explicit form). This
potential is suitable enough for analysis of particle behavior in
wells with enough high barrier (with enough large depth). Models
of one-dimensional motion of particle in such form of potential
have been studied in the literature \cite{Otchik.1996.VANB}.

In the next works it is expected to obtain basic parameters determining
behavior of particle motion in some forms of asymmetric double-well
potentials.

\newpage
\bibliographystyle{h-physrev4}

\bibliography{Disc}

\newpage
\listoffigures

\end{document}